\documentclass[pra,aps,twocolumn,showpacs,superscriptaddress,amsmath,floatfix,amssymb,nofootinbib]{revtex4}

\usepackage{empheq}
\usepackage{graphicx}
\usepackage{bm}
\usepackage{bbold}
\usepackage{color}

\newcommand{\beq}{\begin{equation}}
\newcommand{\eeq}{\end{equation}}
\newcommand{\beqa}{\begin{eqnarray}}
\newcommand{\eeqa}{\end{eqnarray}}
\def\ra{\rangle}
\def\la{\langle}

\begin{document}
\title{Pulse design without rotating wave approximation}
\author{S. Ib\'a\~{n}ez}
\affiliation{Departamento de Qu\'{\i}mica F\'{\i}sica, Universidad del Pa\'{\i}s Vasco UPV/EHU,
Apdo. 644, Bilbao, Spain}
\author{Yi-Chao Li}
\affiliation{Department of Physics, Shanghai University, 200444 Shanghai, People's Republic of China}
\author{Xi Chen}
\affiliation{Department of Physics, Shanghai University, 200444 Shanghai, People's Republic of China}
\author{J. G. Muga}
\affiliation{Departamento de Qu\'{\i}mica F\'{\i}sica, Universidad del Pa\'{\i}s Vasco UPV/EHU,
Apdo. 644, Bilbao, Spain}
\affiliation{Department of Physics, Shanghai University, 200444 Shanghai, People's Republic of China}
\begin{abstract}
We design realizable time-dependent semiclassical pulses  to  
invert the population of a two-level system faster than adiabatically when the rotating-wave approximation cannot be applied. 
Different approaches, based on the counterdiabatic method or on invariants, 
may lead to singularities in the pulse functions. 
Ways to avoid or cancel the singularities are put forward when the pulse spans few oscillations.  
For many oscillations an alternative numerical minimization method is proposed and demonstrated. 
\end{abstract}
\maketitle
\section{Introduction}
Controlling accurately the internal states of quantum two-level systems, realized by real or artificial atoms,
as in crystal defects, quantum dots, or superconducting qubits,
is a fundamental task  in nuclear magnetic resonance, metrology or to develop new quantum technologies \cite{sta,sun,lin,tseng,deschamps,collin,torosov,sillanpaa,leek}.       
Pulse engineering is the art and science of designing realizable control fields to perform specific operations.  
We assume here that the field is intense enough to be treated  semiclassically. 
The task amounts to solving an ``inverse problem'', as the aim is to determine a realizable Hamiltonian
that drives the system to specific states at a given final time, e.g., the ones that would be reached in an ideal adiabatic driving \cite{ChenPRA}.   
The design and implementation of the pulse are easier for adiabatic dynamics, as the final atomic state is quite insensitive to 
smooth deviations from the ideal pulse, but for faster-than-adiabatic processes, the design and its implementation become more demanding.  
In addition, when the rotating-wave approximation (RWA) \cite{Allen} fails for strong drivings,   
the inversion process becomes more involved. However, this regime is key to achieve fast control of two-level systems, e.g., in
nitrogen-vacancy centers \cite{fuchs,london,childress}.   

In this paper we focus on speeding up ``rapid adiabatic passage" (RAP) population inversion processes in two-level systems \cite{ChenPRA, ChenPRL, zhang},
beyond the RWA. 
Without the RWA, a consistency condition between the diagonal and the non-diagonal elements of the interaction-picture Hamiltonian, which involves the phase $\varphi(t)$ of the field and its derivative with respect to time, $\dot{\varphi}(t)$, must be satisfied.   
We shall first apply the counterdiabatic (CD) method \cite{Rice03,Rice05,Rice08,Berry09,ChenPRL}. In this method, a reference Hamiltonian, i.e., the Hamiltonian for the two-level system without the RWA in the interaction picture, $H(t)$, is complemented (or even substituted) by a CD Hamiltonian, $H_1(t)$, so that the 
system follows exactly the adiabatic dynamics of $H(t)$. 
In the dynamics driven by $H(t)+H_1(t)$, the CD term suppresses transitions in the instantaneous eigenbasis of $H$, but allows for transitions in the instantaneous eigenbasis of $H+H_1$. This Hamiltonian, however, 
does not necessarily satisfy the consistency condition mentioned above, unless an appropriate 
rearrangement of terms is performed, and a new unitary transformation between interaction and
Schr\"odinger pictures, different from the one for $H$, is implemented \cite{prl}.   
{In section II we show that} the consistent field implies in general singularities in the Rabi frequency, 
but the field itself is not singular.  

We use as well invariants of motion \cite{LR}. 
Designing the invariant is equivalent to imposing the desired dynamics. This is easy  
within the RWA by setting the time dependence of independent auxiliary variables corresponding to polar, $\theta(t)$, and azimuthal angles, $\beta(t)$, 
that characterize the invariant eigenstates on the Bloch sphere \cite{ChenPRA}.   
From these angles the time dependences of the Hamiltonian components,
the Rabi frequency $\Omega_R(t)$ and the detuning $\Delta(t)$, and 
thus the physical fields may be deduced
\cite{ChenPRA}.  By contrast, without applying the RWA, 
a naive independent design leads to singularities in the field.   
In section III we show two different ways to avoid this problem.   
When the pulse duration spans only a few field oscillations
the angles $\theta$ and $\beta$ may be set to cancel  
all singularities and produce a smooth, finite-intensity pulse.  
Instead, if many field oscillations occur,  a numerical minimization method to find optimal parameters
in a predetermined pulse form may be used to invert the population.  

We shall set first the basic concepts and notation. 
Assuming a semiclassical interaction between the electric field, 
$E(t)=E_0(t)\cos[\varphi(t)]$, and the two-level atom, 
the Hamiltonian of the atom in the Schr\"odinger picture, in the electric dipole approximation, is, see e.g.  \cite{Alvaro},
\beqa
H_{s}(t)&=&  \frac{\hbar}{2} \Big\{ \omega_{0}(t) (|e\rangle \langle e|-|g\rangle \langle g|)
\nonumber\\
&+& \Omega_{R}(t) [(|e\rangle \langle g|+|g\rangle \langle e|) (e^{i\varphi(t)}+e^{-i\varphi(t)})]  \Big\},
\label{hami}
\eeqa
in the bare basis of the atom $|g\ra = (\tiny{\begin{array} {c} 1\\ 0 \end{array}})$, 
$|e\ra =(\tiny{\begin{array} {c} 0\\ 1 \end{array}})$,
where
 $\omega_0(t)/(2\pi)$ is the transition frequency of the atom, which may in general depend on time, controlled, e.g., by Stark shifts;
$\Omega_R(t)$ is the Rabi frequency, assumed real (without loss of generality for transitions 
without change in the magnetic number, see \cite{Allen});
and $\varphi(t)$ is the time dependent phase of the electric field of the pulse,
where $\dot{\varphi}(t)/(2\pi)$ is the instantaneous field frequency. 
The ``exact'' Hamiltonian, i.e., without applying the RWA, in a 
field-adapted interaction picture is given by \cite{Alvaro}
\beq
H(t) =  U^{\dagger}_{\varphi} (H_s-H_{\varphi}) U_{\varphi},
\label{trans}
\eeq
where
\beq
H_{\varphi}(t)=\frac{\hbar\dot{\varphi}}{2} (|e\ra\la e|-|g\ra\la g|)
\label{H_phi}
\eeq
and 
\beqa
U_{\varphi}(t) &=& e^{-i\int_0^t{H_{\varphi}(t') dt'}/\hbar}
\nonumber\\
&=& e^{-i\varphi(t)/2} |e \rangle \langle e|+e^{i\varphi(t)/2} |g\rangle \langle g|
\label{unit}
\eeqa
is the unitary operator of the transformation. Thus,
\beqa
H(t)= \frac{\hbar}{2}
\left(
\begin{array}{cc}
-\Delta& \Omega
\\
\Omega^* & \Delta
\end{array}
\right),
\label{H}
\eeqa
with
\beq
\Omega(t)=  \Omega_R(t)  [1+e^{-2i \varphi(t)}],
\label{Omega}
\eeq
and detuning  
\beq
\Delta(t)= \omega_0(t)-\dot{\varphi}(t).
\label{delta}
\eeq
The eigenenergies of  $H$ are $E_{\pm}(t) =\pm \hbar \varepsilon_0(t)/2$, with
$\varepsilon_0(t)= \sqrt{\Delta(t)^2+|\Omega(t)|^2}$, and the  (time dependent) eigenstates,
$|{\pm}\ra \equiv |{\pm}(t)\ra$, are
\beqa
|\pm\ra &=& \left[\frac{-(\Delta\mp \varepsilon_0)} {\Omega^*} |g\ra
+  |e\ra \right] \frac{1} {\sqrt{1+ \frac{(\Delta\mp \varepsilon_0)^2}{|\Omega|^2}}},
\label{vectors}
\eeqa
which satisfy $H(t) |{\pm}(t)\ra = E_{\pm}(t) |{\pm}(t)\ra$.

The exact two-level system Hamiltonian in the interaction picture entails $\dot{\varphi}(t)$ in the detuning, i.e., in the diagonal elements, see Eq. (\ref{delta}), and its integral, $\varphi(t)$, in the non-diagonal elements, {see Eq. (\ref{Omega})}.
By ``consistency condition'' we mean that the elements of a physically allowed 
interaction picture Hamiltonian must comply with the structure 
set in Eqs. (\ref{Omega}) and (\ref{delta}).  
In particular, this structure must be satisfied  when
designing pulses to carry out faster-than-adiabatic inversion processes, as we shall see in sections II and III.
Suppose for example that the functions $\Omega_R(t)$ and $\omega_0(t)$ are given. Then, not any Hamiltonian is allowed (consistent) since the diagonal and 
non-diagonal parts must depend on $\varphi(t)$ and its derivative consistently.   
\section{The counterdiabatic method}
The counterdiabatic approach adds a counterdiabatic or ``CD'' term to some reference Hamiltonian to make the exact  
dynamics adiabatic with respect to the reference Hamiltonian \cite{Rice03,Rice05,Rice08,Berry09,ChenPRL}. 
The formal construction of the CD term is explained in the original references  \cite{Rice03,Rice05,Rice08,Berry09,ChenPRL,ChenPRA}. 
We shall apply the counterdiabatic method to speed up an adiabatic population inversion process for a two-level systems beyond the RWA, where the (reference) Hamiltonian of the system is given by Eq. (\ref{H}), or in diagonal form as
$H(t)=\sum_n |n(t)\ra E_n(t)\la n(t)|$, where $|n(t)\ra = |\pm(t)\ra$. 
The inversion is from  
$|\Psi(0)\ra = |g\ra$ to $|\Psi(t_f)\ra = |e\ra$, up to a phase factor, where $t=0$ and $t=t_f$ are the initial and final times of the process, and $|\Psi(t)\ra$ is the general state of the system.
We consider a constant field (angular) frequency,
$\dot{\varphi}(t) = \omega_L$, so that $\varphi(t)= \omega_L t$, and a time dependent transition frequency,
$\omega_0(t)/(2\pi)$, that will determine the  detuning. Then, from Eqs. (\ref{Omega}) and (\ref{delta}),
\beqa
\Omega(t) &=& \Omega_R (1+ e^{-2i \omega_L t}),
\label{Omega0} \\
\Delta(t) &=& \omega_0(t)-\omega_L.
\label{delta1}
\eeqa
The CD term is in general given by, see e.g.  \cite{Berry09},
\beqa
H_1 (t) &=& i\hbar \sum_n (|\partial_t{n}\ra \la n| - \la n|\partial_t{n}\ra |n\ra \la n|)
\nonumber \\
&=& i \hbar \sum_{m \neq n} \sum \frac{|m\ra \la m| \partial_t H |n\ra \la n|}{E_n -E_m}.
\nonumber
\eeqa
For the {exact} two-level system, using Eq. (\ref{vectors}), $H_1$ is given by \cite{ChenWei}
\beqa
H_1 (t)&=&\frac{i \hbar}{2 C_1}
\left(
\begin{array}{cc}
- B_1/2 & A_1
\\
-A_1^* & B_1/2
\end{array}
\right), 
\nonumber
\eeqa
where $A_1(t)=  \dot{\Omega} \Delta -\dot{\Delta} \Omega$,
$B_1(t)=  \dot{\Omega}^*  \Omega - \dot{\Omega} \Omega^*$ is purely imaginary, and
$C_1(t)= \Delta^2 + |\Omega|^2$.
Thus, the total Hamiltonian provided by the CD method is 
\beqa
H+H_1 &=& \frac{\hbar}{2}
\left(
\begin{array}{cc}
- \tilde{\Delta} & \tilde{\Omega}
\\
\tilde{\Omega}^* & \tilde{\Delta}
\end{array}
\right),
\label{H0H1new}
\eeqa
where
\beqa
\tilde{\Omega}&=& \Omega + iA_1/C_1,
\label{Omega_tilde}
\\
\tilde{\Delta}&=&\Delta + iB_1/2C_1.
\label{detutot}
\eeqa
From Eq. (\ref{delta1}), the detuning of the total Hamiltonian, given by Eq. (\ref{detutot}), becomes
\beq
\tilde{\Delta} = (\omega_0+ iB_1/2C_1)-\omega_L.
\label{delta_tilde}
\eeq
The term $\omega_0 + iB_1/2C_1$ may be interpreted as a new time-dependent {transition (angular)} frequency, and $\omega_L/(2\pi)$ is, as in the reference Hamiltonian, the  
constant field frequency. 
Then, the non-diagonal element $\tilde{\Omega}$, given by Eq. (\ref{Omega_tilde}), should be expressed as
$\tilde{\Omega}= \tilde{\Omega}_R  (1+e^{-2i \omega_L t})$, compare with Eq. (\ref{Omega0}), in order to satisfy the consistency condition, where $\tilde{\Omega}_R \equiv \tilde{\Omega}_R(t)$ is a new Rabi frequency corresponding to the pulse associated with
$H+H_1$.

\begin{figure}[t]
\begin{center}
\includegraphics[height=4.0cm,angle=0]{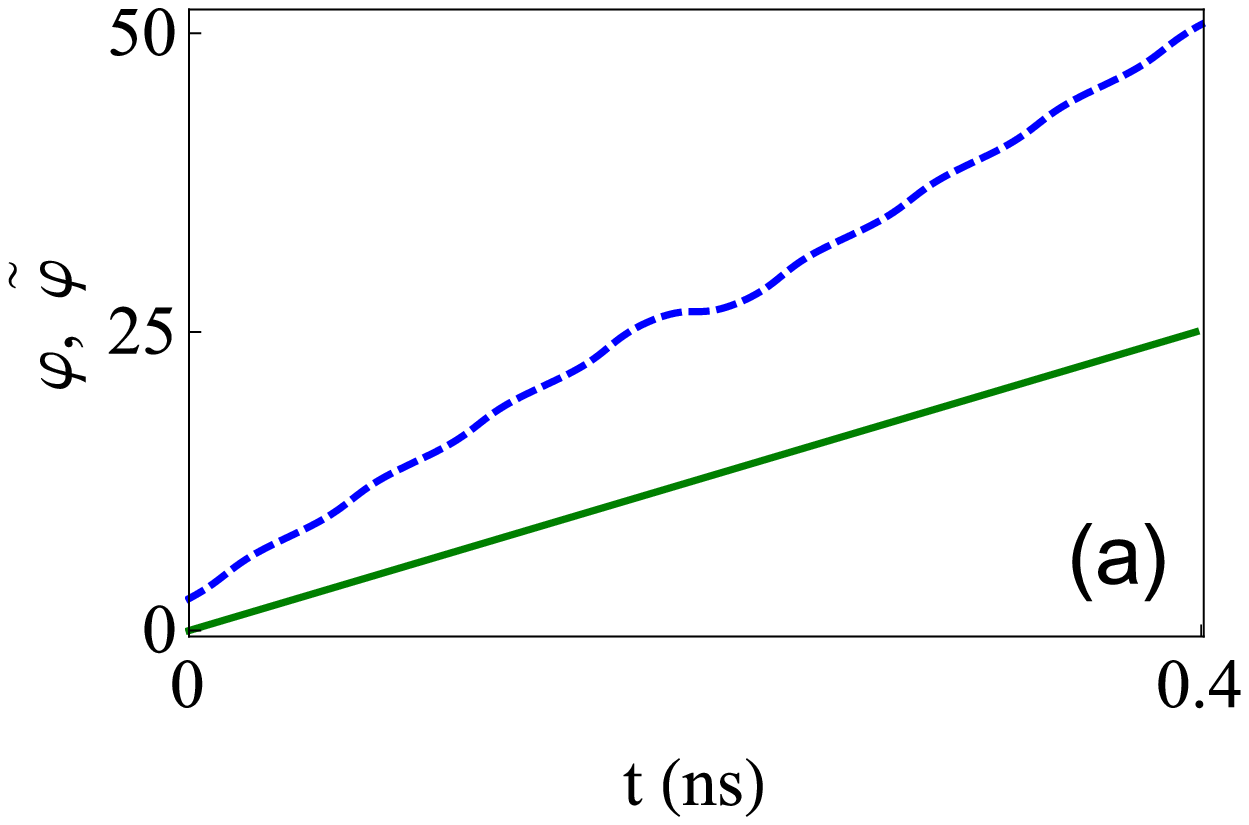}
\includegraphics[height=4.0cm,angle=0]{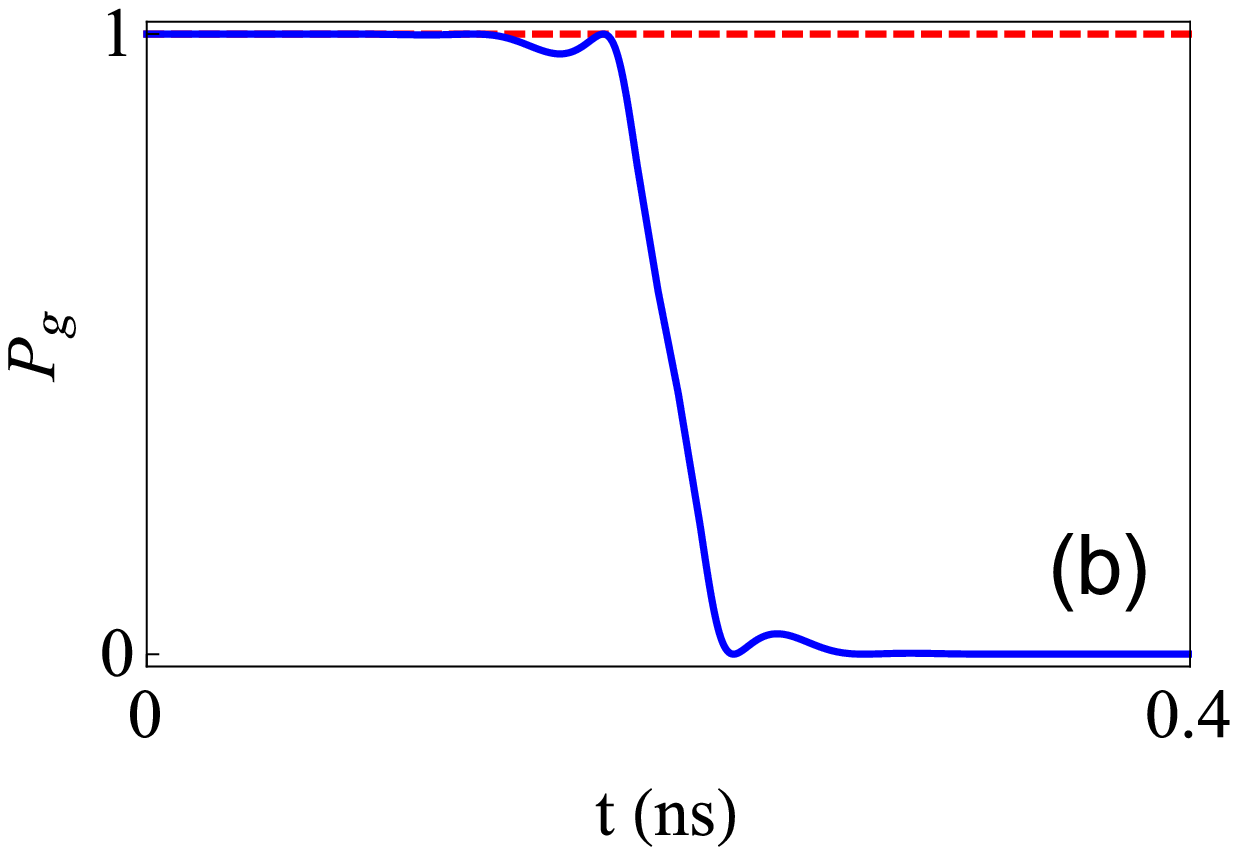}
\includegraphics[height=4.0cm,angle=0]{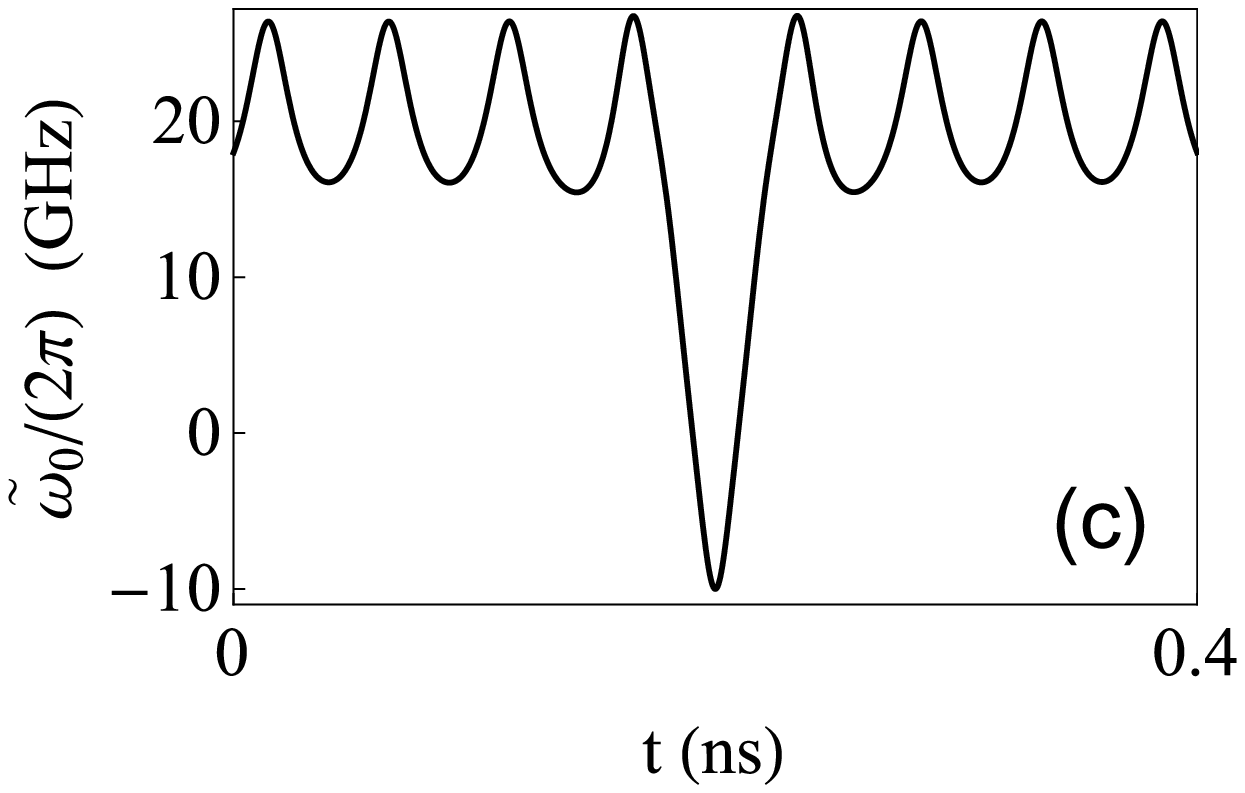}
\end{center}
\caption{\label{chen_wei}
(Color online)
(a) $\varphi(t)= \omega_L t$ (green solid line) and $\tilde{\varphi}(t)$ {given by Eq. (\ref{argument})}
(blue short-dashed line),
(b) $P_g$ driven by $H$ (red dashed line) and by $H + H_1$ (blue solid line), and
(c) $\tilde{\omega}_0(t)= \tilde{\Delta} + \dot{\tilde{\varphi}}$ {divided by $2\pi$},
for the Allen-Eberly protocol and parameters:
$\Omega_M= 2\pi \times 3$ MHz, $\delta= 2\pi \times 200$ MHz, $t_0= 0.05$ ns, $\omega_L= 2\pi \times 10$ GHz,
and $t_f= 0.4$ ns.
}
\end{figure}

\begin{figure}[t]
\begin{center}
\includegraphics[height=5.5cm,angle=0]{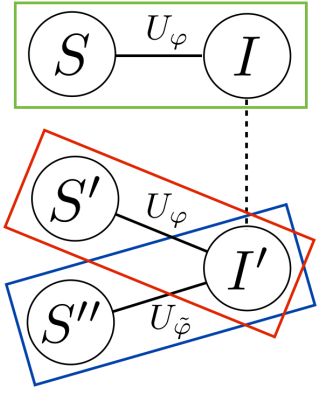}
\end{center}
\caption{\label{figure}
(Color online)
Schematic relation between different Schr\"{o}dinger and interaction picture dynamical equations. Each node may represent the dynamical equations, and also the Hamiltonian or the states. The rectangular boxes enclose nodes that represent the same underlying physics. The solid lines represent unitary relations for the linked states and the dashed line represents a non-unitary addition of a term to the Hamiltonian.}
\end{figure}

To see whether $H+H_1$ is indeed consistent 
we first rewrite the complex $\tilde{\Omega}$ in Eq. (\ref{Omega_tilde}) in a convenient form,  
\beqa
\tilde{\Omega} &=&\tilde{\Omega}_R  (1+e^{-2i \tilde{\varphi}})
\label{form}
\\
&=& 2\tilde{\Omega}_R \cos{\tilde{\varphi}} e^{-i \tilde{\varphi}}, 
\label{Omega1}
\eeqa
where $\tilde{\Omega}_R$ and $\tilde{\varphi} \equiv \tilde{\varphi}(t)$ are real.   
$\tilde{\varphi}$ is given by 
\beq
\tilde{\varphi} = -\arg(\tilde{\Omega})+2\pi n, 
\label{argument}
\eeq
where n is an integer chosen to make $\tilde{\varphi}$ continuous. 
Once $\tilde{\varphi}$ is determined, taking into account Eq. (\ref{Omega_tilde}),
$\tilde{\Omega}_R$ is calculated from Eqs. (\ref{form}) or (\ref{Omega1}).  
However, the consistency condition is generally not satisfied, i.e.,
$\tilde{\varphi}\neq \varphi=\omega_L t $.
For the Allen-Eberly protocol \cite{Allen},
\beqa
\Omega_R(t) &=& \Omega_M \sinh{\left[\frac{\pi (t-t_f/2)} {2 t_0}\right]},
\nonumber \\
\Delta(t) &=& \frac{2 \delta^2 t_0}{\pi} \tanh{\left[\frac{\pi (t-t_f/2)} {2 t_0}\right]},
\label{delta_allen}
\eeqa
and parameters $\Omega_M= 2\pi \times 3$ MHz, $\delta= 2\pi \times 200$ MHz, $t_0= 0.05$ ns,
$\omega_L= 2\pi \times 10$ GHz, and $t_f= 0.4$ ns,
Fig. \ref{chen_wei} (a) shows that $\varphi(t)= \omega_L t$ and $\tilde{\varphi}(t)$ do not coincide.
These parameters are chosen so that $H(t)$ does not invert the populations of the bare basis,
$P_g(t)= |\la g|\Psi(t) \ra|^2$ and $P_e(t)= |\la e|\Psi(t) \ra|^2$.
Fig. \ref{chen_wei} (b) shows that the Hamiltonian $H+H_1$ does invert the population. The 
CD term works formally, but the Hamiltonian $H+H_1$ does not correspond to a field 
with frequency $\omega_L/(2\pi)$.
If we apply the same transformation that relates the Schr\"odinger picture Hamiltonian $H_s$ and {the interaction-picture Hamiltonian} $H$,  
\beqa
U_{\varphi}(t) = e^{-i\omega_L t/2} |e\rangle \langle e|+e^{i\omega_L t/2} |g\rangle \langle g|,
\label{unit1}
\eeqa
for $\varphi= \omega_L t$, {following Eq. (\ref{trans})},
the Schr\"odinger picture Hamiltonian corresponding to $H+H_1$ does not take the 
form of Eq. (\ref{hami}) with modified functions for the transition and Rabi frequencies.  
This procedure is schematized in Fig. \ref{figure} by the boxes around $S,\,I$, and $S',\, I'$,
where $S$ represents the initial
Schr\"{o}dinger picture driven by $H_s$ and $I$ the interaction picture driven by $H$ 
with $U_{\varphi}$, given by Eq. (\ref{unit1}), the unitary operator that generates the transformation. $I'$ represents the interaction picture with the addition of an extra term, $H+H_1$, and $S'$ the corresponding Schr\"{o}dinger picture mediated again by Eq. (\ref{unit1}).
%
%
%
%
%
%
%
%
%
%
%
%

We might as well assume that $H_1$ is a Hamiltonian corresponding to an independent (second) pulse complementing the pulse for $H$. This, however, leads to the same result as interpreting $H+H_1$ as a single pulse, since $H$ is negligible versus $H_1$, i.e., 
$|\Omega| \ll |iA_1/C_1|$  and $\Delta \ll iB_1/(2 C_1)$, for the given parameters. 
The consistency condition is again not satisfied.

\begin{figure}[t]
\begin{center}
\includegraphics[height=4.5cm,angle=0]{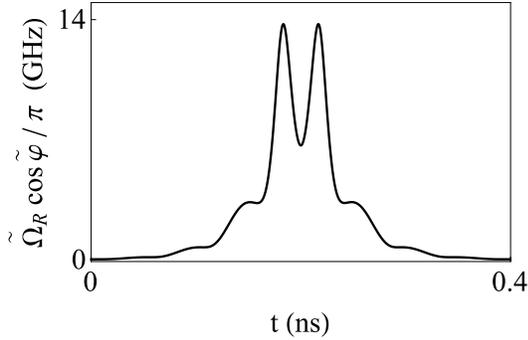}
\end{center}
\caption{\label{field}
$2 \tilde{\Omega}_R \cos{\tilde{\varphi}} /(2\pi)$ 
for parameters as in the caption of Fig. \ref{chen_wei}.
}
\end{figure}

To enforce the consistency condition between the diagonal and non-diagonal elements in Eq. (\ref{H0H1new}) we may 
rewrite $\tilde{\Delta}$, see Eq. (\ref{delta_tilde}),  with the required structure, i.e.,  
$\tilde{\Delta}= \tilde{\omega}_0(t)-\dot{\tilde{\varphi}}$, where $\tilde{\omega}_0(t)/(2\pi)$ is a new time-dependent 
transition frequency of the atom. Equating this expression to Eq. (\ref{delta_tilde}) gives  
\beq
\tilde{\omega}_0(t) = \omega_0-\omega_L+ iB_1/2C_1 + \dot{\tilde{\varphi}},
\nonumber
\eeq
which is depicted in Fig. \ref{chen_wei} (c) with $\omega_0(t)$ obtained from Eq. (\ref{delta1}) and $\Delta$ given by Eq. (\ref{delta_allen}). 
The unitary  transformation $U_{\tilde{\varphi}}$, where $\varphi$ is substituted by $\tilde{\varphi}$ in Eqs.  (\ref{H_phi}) and (\ref{unit}),
leads to a different Schr\"{o}dinger picture $S''$ \cite{prl}, see Fig. \ref{figure}.
The Hamiltonian in the Schr\"{o}dinger picture $S''$ is
\beqa
H_{S''}(t) &=& \frac{\hbar}{2}
\left(
\begin{array}{cc}
-\tilde{\omega}_0&2\tilde{\Omega}_R\cos\tilde{\varphi} 
\\
2\tilde{\Omega}_R\cos\tilde{\varphi} & \tilde{\omega}_0
\end{array}
\right).
\nonumber
\eeqa
This procedure solves the consistency condition, 
but the new Rabi frequency, {given  by Eqs. (\ref{form}) or (\ref{Omega1})}, 
is singular at the zeros of $\cos{\tilde{\varphi}}$. 
However, the field,
proportional to $\tilde{\Omega}_R\cos{\tilde{\varphi}}$, is well behaved, see Fig. \ref{field}.
%
%
%
%
%
%
%

In the following section we work out a different approach based on invariants of the motion.

\section{Invariant-based inverse engineering}
Associated with $H(t)$ there are hermitian dynamical invariants, $I(t)$, satisfying  the invariance condition \cite{LR},
\beqa
\frac{\partial{I}}{\partial{t}}+\frac{1}{i\hbar} [I,H] = 0,
\label{cond}
\eeqa
that may be parametrized as \cite{ChenPRA,inva,inva1}
\beqa
I(t)=\frac{\hbar I_0}{2}
\left(
\begin{array}{cc}
\cos{\theta} & \sin{\theta} e^{-i\beta}
\\
\sin{\theta} e^{i\beta} & -\cos{\theta}
\end{array}
\right),
\label{invariant}
\eeqa
where $I_0$ is an arbitrary constant angular frequency to keep $I(t)$ with dimensions of energy, and
$\theta \equiv \theta(t)$ and $\beta \equiv \beta(t)$ are the polar and azimuthal angles in the Bloch sphere, respectively. 
The eigenvalue equation for $I(t)$ is $I(t) |\phi_{\pm}^I(t)\ra = \lambda_{\pm} |\phi_{\pm}^I(t)\ra$, where the eigenvalues are
$\lambda_{\pm}= \pm \hbar I_0/2$ and, consistently with orthogonality and normalization, we can choose the basis of eigenstates 
\beq
|\phi_+^I(t)\ra=  \Bigg(\!\begin{array} {c} \cos{\frac{\theta}{2}} e^{-i\frac{\beta}{2}}\\ \sin{\frac{\theta}{2}} e^{i\frac{\beta}{2}} \end{array}\!\Bigg), \,
|\phi_-^I(t)\ra=  \Bigg(\!\begin{array} {c} \sin{\frac{\theta}{2}} e^{-i\frac{\beta}{2}}\\ -\cos{\frac{\theta}{2}} e^{i\frac{\beta}{2}} \end{array}\!\Bigg).
\label{eigenstates_of_I}
\eeq
The solution of the time-dependent Schr\"{o}dinger equation,
$i\hbar \partial_t |\Psi(t)\ra = H(t) |\Psi(t)\ra$, can be expressed as \cite{LR}
\beq
|\Psi(t)\ra=  c_+ e^{i \gamma_+(t)} |\phi_+^I(t)\ra + c_- e^{i \gamma_-(t)} |\phi_-^I(t)\ra,
\nonumber
\eeq
up to a global phase factor,
where the $c_{\pm}$ are time-independent coefficients and the $\gamma_{\pm}(t)$ are the Lewis-Riesenfeld phases,
\beq
\gamma_{\pm}(t) =
\frac{1}{\hbar} \int_{0}^t  \Big\la  \phi_{\pm}^I(t')\Big|i\hbar\frac{\partial}{\partial t'}-H(t')\Big|\phi_{\pm}^I(t') \Big\ra dt'.
\nonumber
\eeq
The Lewis-Riesenfeld phase becomes a global phase if the dynamics is carried out by one eigenstate of the invariant only. 
This will be the case in the inversions discussed here and leads to an important simplification: $\gamma_{\pm}(t)$ can be ignored
to engineer  the Hamiltonian.    

From the invariance condition (\ref{cond}), with $I$ given by Eq. (\ref{invariant}) and $H$ given by Eq. (\ref{H}), we find 
\beqa
\dot{\theta} &=& -2\Omega_R \cos\varphi\sin(\beta-\varphi),
\nonumber\\
\dot{\beta} &=& - \Delta -2 \Omega_R\cot{\theta} \cos{\varphi} \cos(\beta-\varphi).
\label{NRWA}
\eeqa
The first equation bounds $\dot{\theta}$ between $-2\Omega_R$ and $2\Omega_R$.
These equations 
are much simplified when the RWA may be applied \cite{ChenPRA}. 
The two-level system Hamiltonian under the RWA is, see for example  \cite{ChenPRA},
\beqa
H_{RWA}(t)=\frac{\hbar}{2}
\left(
\begin{array}{cc}
-\Delta & \Omega_R 
\\
\Omega_R & \Delta
\end{array}
\right), 
\nonumber
\eeqa
where $\varphi(t)$ is absent in the non-diagonal elements of $H_{RWA}(t)$, 
compare with $H(t)$ in Eq. (\ref{H}). 
Now the spherical angles satisfy  \cite{ChenPRA,inva,inva1}
\beqa
RWA\left\{
\begin{array}{lll}
\dot{\theta} &=& - \Omega_R\sin{\beta},
\\
\dot{\beta} &=& -\Delta - \Omega_R\cot{\theta}\cos{\beta}.
\end{array}
\label{RWA}
\right.
\eeqa

The ``direct problem" is to solve the systems of differential equations (\ref{NRWA}) or (\ref{RWA}) 
for $\theta$ and $\beta$ when $\Omega_R$ and $\Delta$ are given, once we fix $\omega_0$ or $\varphi$ for the system (\ref{NRWA}),
see also Eq. (\ref{delta}).  
Instead, in the ``inverse problem'' we have in principle  to construct the  functions
$\Omega_R$ and $\Delta$ from 
$\theta$ and $\beta$. 
When the RWA holds, from the system (\ref{RWA}),
the inversion reduces to 
simple expressions for $\Omega_R$ and $\Delta$ \cite{ChenPRA},
\beqa
RWA\left\{
\begin{array}{lll}
\Omega_R&=& -\dot{\theta} / \sin\beta,
\\
\Delta&=& - \dot{\beta} + \dot\theta\cot\theta\cot\beta.
\end{array}
\nonumber
\right.
\eeqa
Without  the RWA, from the system (\ref{NRWA}), taking into account Eq. (\ref{delta}), we get 
\beqa
\Omega_R&=& - \frac{\dot{\theta}}{2\cos\varphi \sin{\alpha}},
\label{OmegaR}
\\
\dot{\alpha}&=& -\omega_0 + {\dot{\theta}}{\cot{\theta}} \cot{\alpha},
\label{system}
\eeqa
where $\alpha(t)=\beta(t)-\varphi(t)$ is introduced to simplify the expressions.     
If the inversion strategy is to consider $\omega_0(t)$ as given, which in particular could be constant, and  $\theta(t)$
is designed,  
the  differential equation for $\alpha(t)$, see Eq. (\ref{system}), is problematic, as an arbitrary choice of $\theta(t)$ and
$\omega_0(t)$ will typically lead to 
singularities on the right hand side, and $\alpha(t)$ will, in general, introduce singularities in $\Omega_R(t)$.
Eq. (\ref{OmegaR}) makes clear that a finite, smooth $\Omega_R$ and a phase 
$\varphi$ increasing along many field cycles, 
require many zeros of $\dot{\theta}$ to compensate for zeros 
in the denominator.  A naive choice for $\theta$ is thus doomed to fail. A different approach is needed.  
It is useful to rewrite Eq. (\ref{system}), taking into account Eq. (\ref{delta}),  as  
\beqa
\Delta= -(\dot{\varphi}+\dot{\alpha}) + {\dot{\theta}}\, {\cot{\theta}} \cot{\alpha}.
\label{detuning}
\eeqa
Even if $\theta$ and $\alpha$ avoid the singularities for a given $\varphi$ in Eq. (\ref{detuning}), 
the required consistency between Eqs. (\ref{delta}) and (\ref{detuning}) (they must be equal)
will in general fail if $\omega_0(t)$ is also given. 
Thus, the proposed strategy is to fix first $\varphi(t)$, then design $\theta(t)$ to avoid the zeros of 
$\cos\varphi$ in Eq. (\ref{OmegaR}), and from there design $\alpha(t)$ to compensate singularities of $\cot\theta$ in Eq. (\ref{detuning}).  This produces a smooth $\Omega_R$
given by Eq. (\ref{OmegaR}), 
and $\Delta$ follows from Eq. (\ref{detuning}). 
Finally, $\omega_0(t)$ is deduced consistently from Eq. (\ref{delta}). 

%
%
%
%
\begin{figure}[t]
\begin{center}
\includegraphics[height=2.8cm,angle=0]{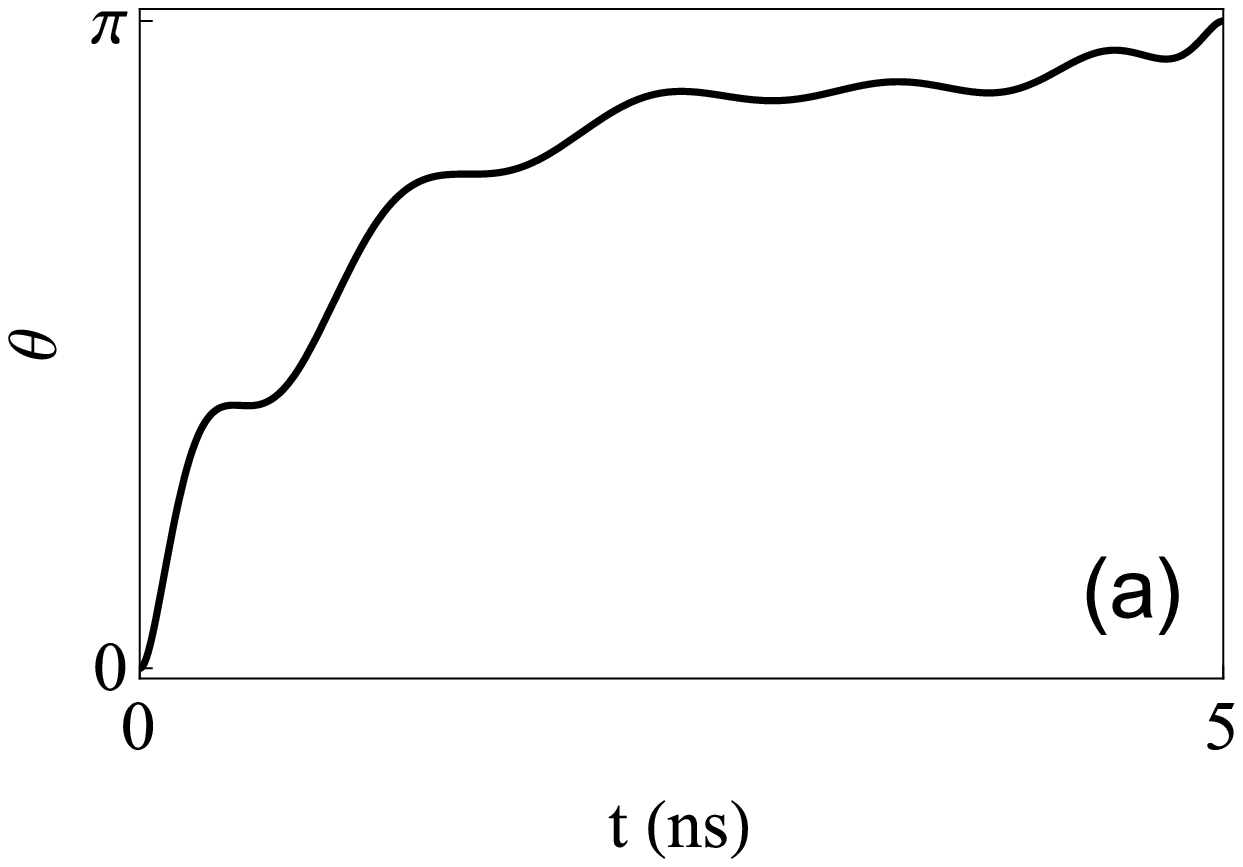}
\includegraphics[height=2.8cm,angle=0]{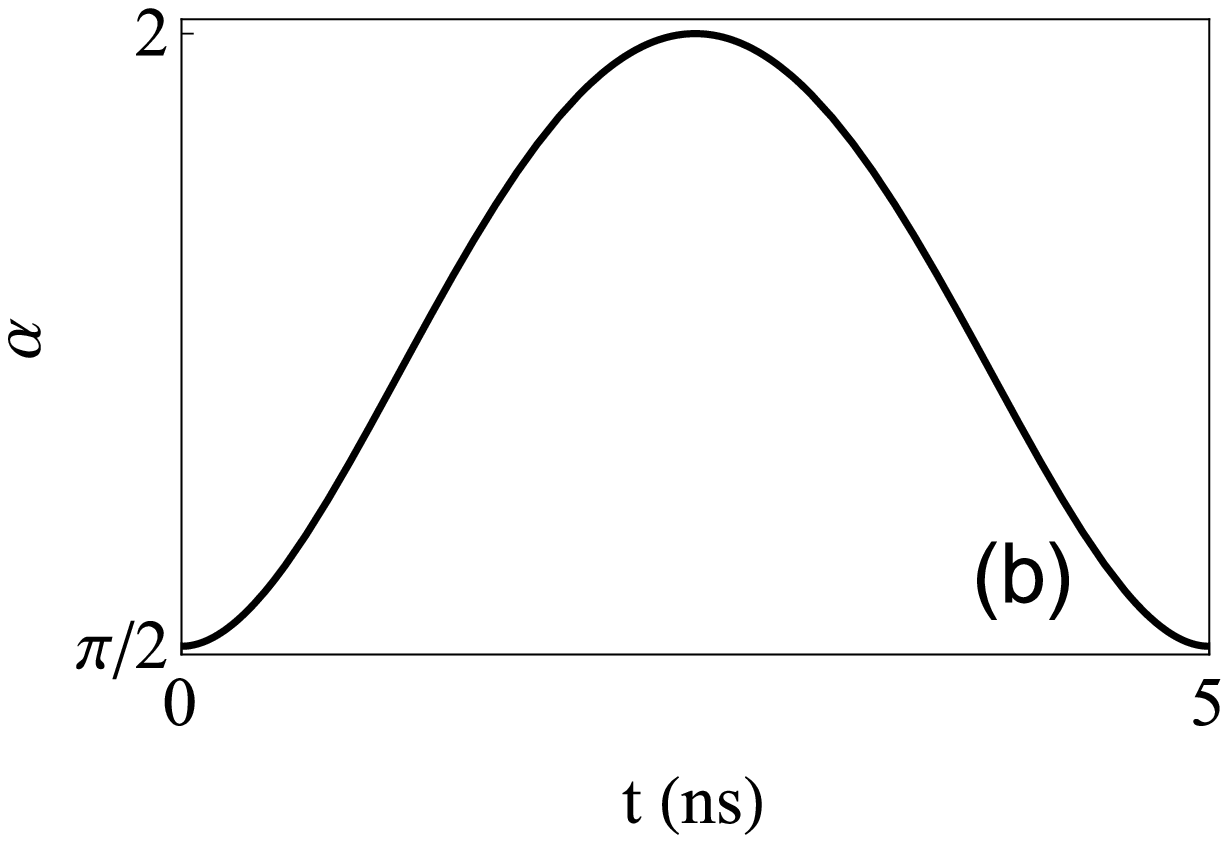}
\includegraphics[height=2.8cm,angle=0]{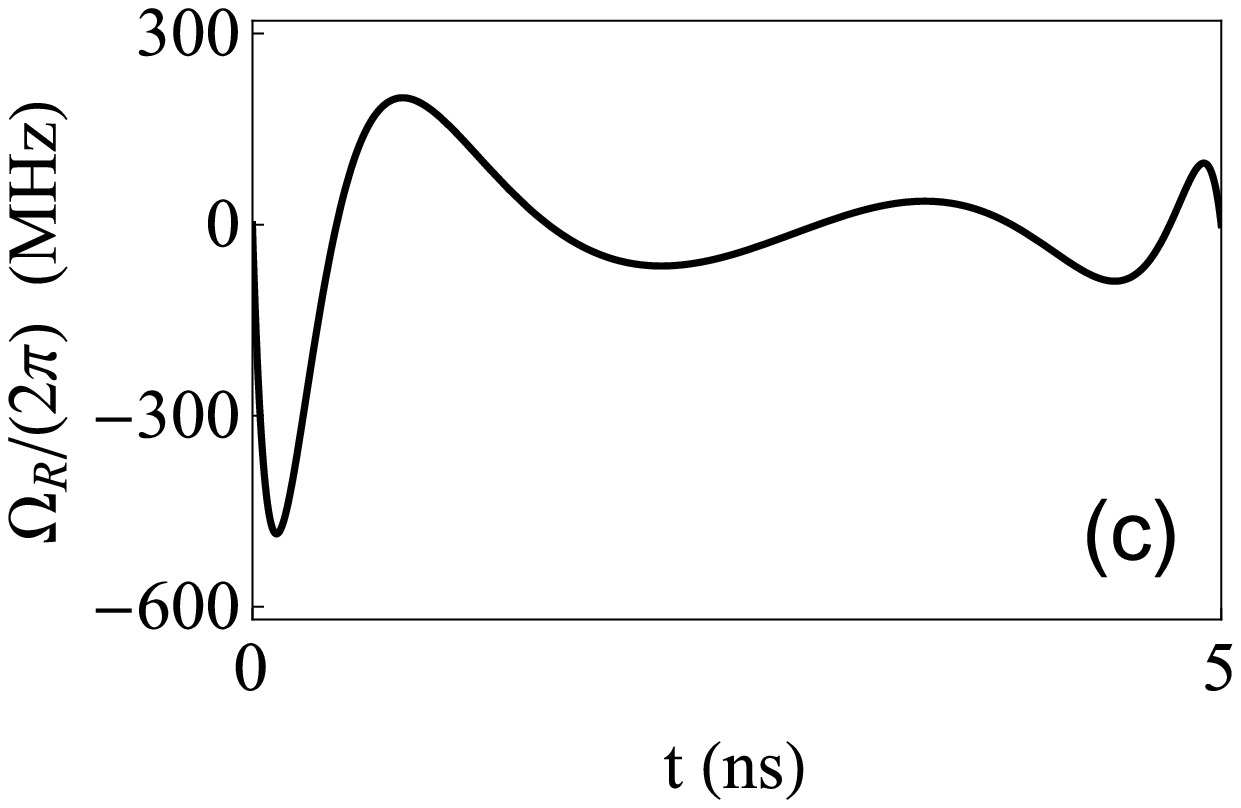}
\includegraphics[height=2.7cm,angle=0]{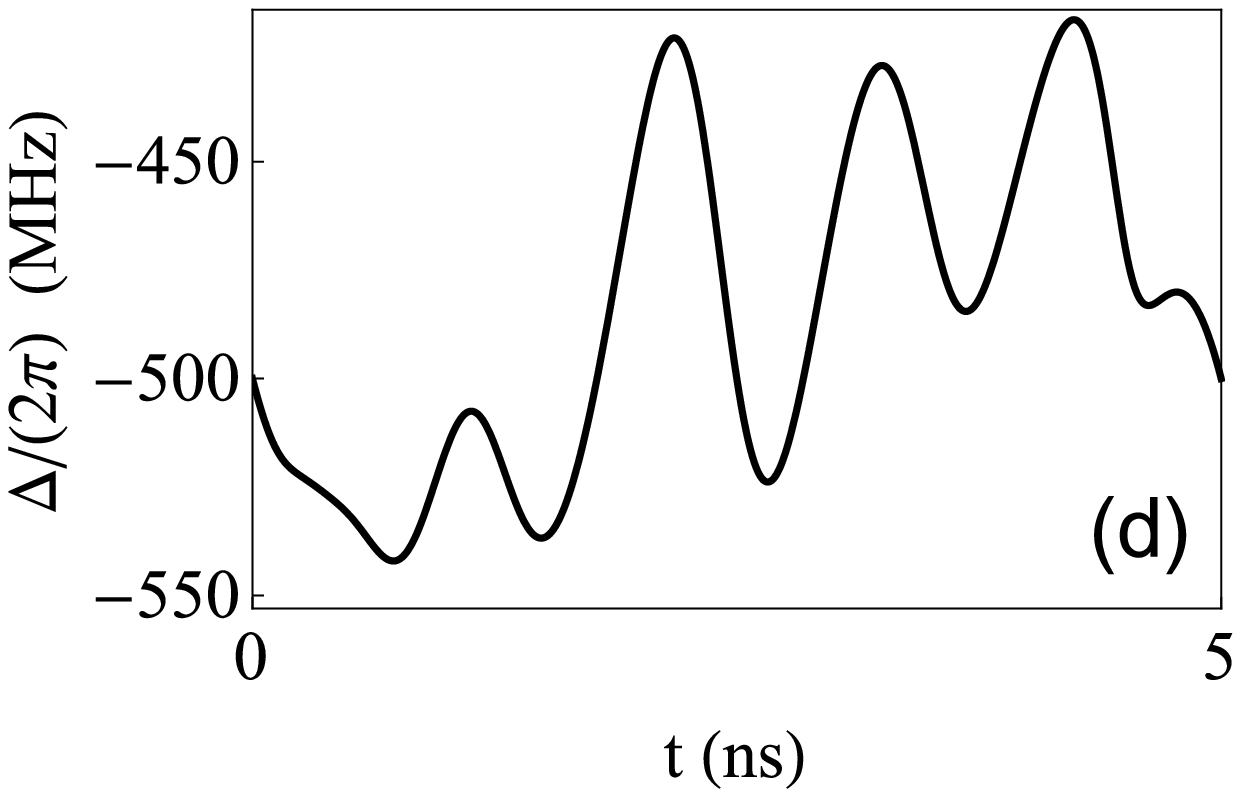}
\includegraphics[height=2.8cm,angle=0]{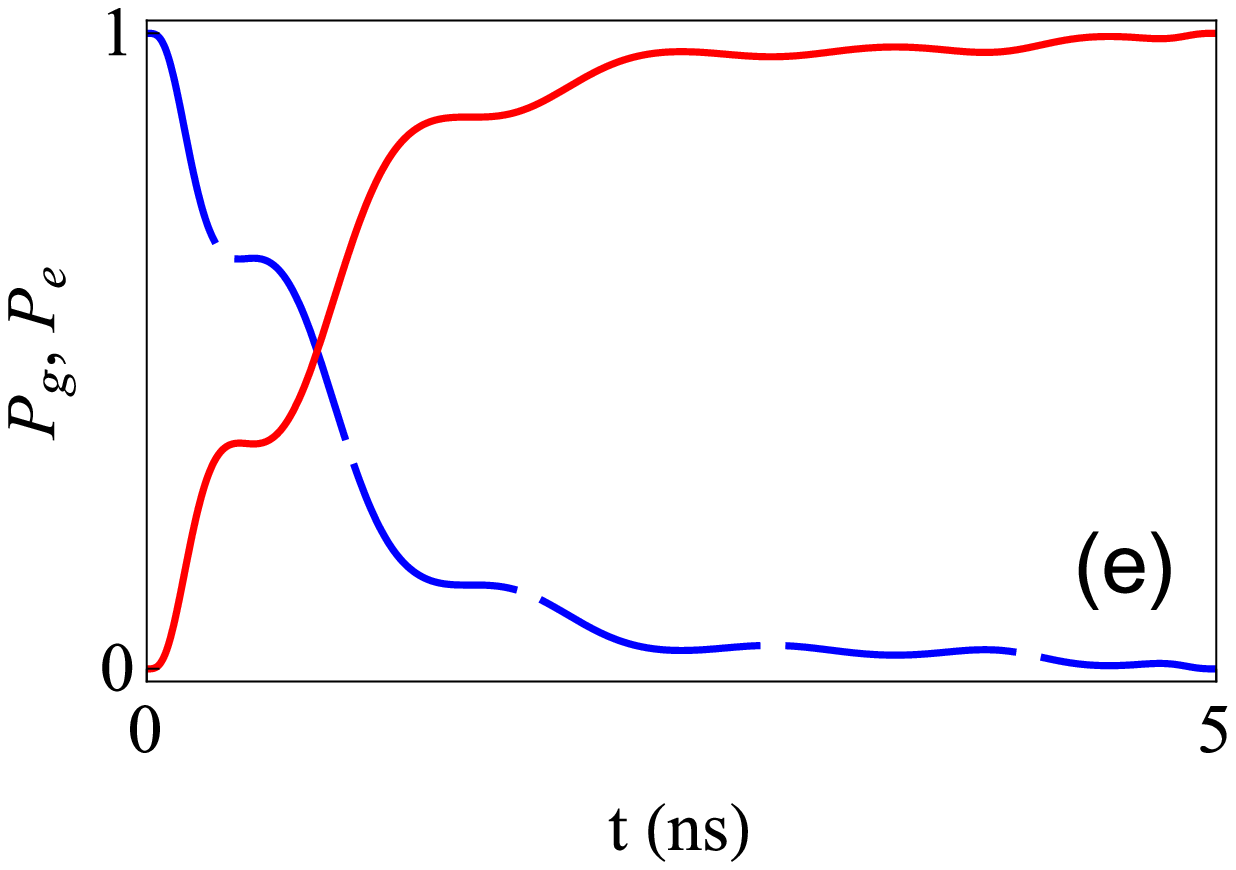}
\includegraphics[height=2.8cm,angle=0]{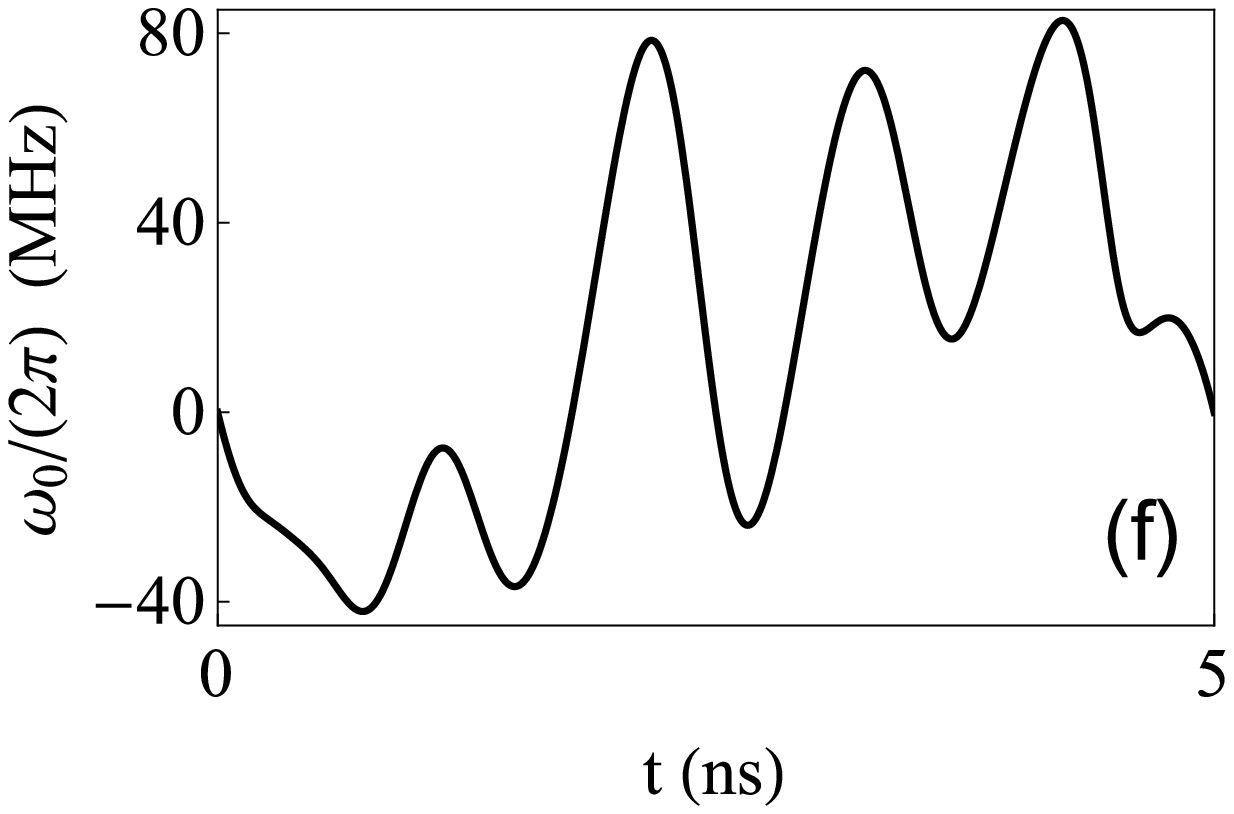}
\end{center}
\caption{\label{few_osci}
(Color online)
Interpolated functions (a) $\theta$ and (b) $\alpha$ for a population inversion process with parameters
$\omega_L= 2 \pi \times 500$ MHz and $t_f= 5$ ns. (c) and (d) show the corresponding
$\Omega_R/(2\pi)$, given by Eq. (\ref{OmegaR}), and $\Delta/(2\pi)$, given by Eq. (\ref{detuning}).
(e) shows the populations of the bare basis, $P_g$ (blue long-dashed line) and $P_e$ (red solid line), for these Hamiltonian functions, and (f) shows the time-dependent atom frequency, $\omega_0/(2\pi)$, derived from $\Delta$ in Eq. (\ref{delta}) as
$\omega_0(t) = \Delta(t) + \omega_L$.} 
\end{figure}
%
%
%
For a population inversion  from 
$|\Psi(0)\ra = |g\ra$ to $|\Psi(t_f)\ra = |e\ra$, up to a phase factor,  we may use $|\phi_+^I(t)\ra$ to carry the dynamics, see
Eq. (\ref{eigenstates_of_I}), with $\theta(t)$ going from $0$ to $\pi$. 
In addition, $H(t)$ and $I(t)$ should commute at 
$t=0$ and $t=t_f$ so that the initial and final states, $|\phi_+^I(0)\ra$ and $|\phi_+^I(t_f)\ra$, are eigenstates 
of the initial and final Hamiltonians. This implies vanishing derivatives of $\theta(t)$ at the initial and final times.  
We have in summary the boundary conditions 
\beqa
\theta(0)=\dot{\theta}(0)=\dot{\theta}(t_f)=0,   \,\,\,\,\,\,\,\,\,\,  \theta(t_f)=\pi. 
\label{BC1}
\eeqa
\subsection{Pulse with few field oscillations}
For pulses containing few field oscillations and assuming again a constant (angular) frequency for the pulse field, 
$\dot{\varphi}= \omega_L$, and $\varphi(t)= \omega_L t$, we may construct $\theta$ so that $\dot{\theta}$ cancels the zeros of $\cos{\varphi}$
in Eq. (\ref{OmegaR}), and then $\alpha$ to compensate the singularities of $\cot{\theta}$ in Eq. (\ref{detuning}),
bounded by $0 < \alpha < \pi$ not to introduce new singularities in
Eqs. (\ref{OmegaR}) and (\ref{detuning}).
In the example of Fig. \ref{few_osci} we use  $\omega_L= 2 \pi \times 500$ MHz and $t_f= 5$ ns, and interpolate
$\theta(t)$ and $\alpha(t)$ with polynomials 
$\theta= \sum_{n=0}^{13} a_n t^n$ and $\alpha= \sum_{n=0}^4 b_n t^n$.
$\theta$ is constructed
to satisfy the boundary conditions (\ref{BC1}), and
to make $\dot{\theta}$ zero at the five intermediate zeros of $\cos{\varphi}$. 
$\theta(1\, {\rm{ns}})=2$,
$\theta(1.6\, {\rm{ns}})=2.4$,
$\theta(2.5 {\rm{ns}})=2.8$,
$\theta(4\, {\rm{ns}})=2.8$, and
$\theta(4.5\, {\rm{ns}})=3$
are also imposed to force a smooth ascent of $\theta$ up to $\pi$,
see Fig. \ref{few_osci} (a).
At the boundary times, $t_b=0, t_f$, the conditions given by Eq. (\ref{BC1}) imply that
$\lim_{t\to t_b} (\dot{\theta} \cot\theta \cot\alpha)=-2\dot{\alpha}(t_b)$, see Eq. (\ref{detuning}).  
The polynomial ansatz for $\alpha$ is constructed
so that $\alpha$ becomes $\pi/2$ 
at the singularities of $\cot{\theta}$,
in this case at $t=0$ and $t=t_f$.
At the boundary times we choose as an example $\dot{\alpha}(0)=\dot{\alpha}(t_f)= 0$, which corresponds to $\omega_0(0)=\omega_0(t_f) =0$.
The  function is additionally tamed for smoothness 
by imposing $\alpha(t_f/2)= 2$, see Fig. \ref{few_osci} (b).
$\Omega_R$ and $\Delta$ calculated from Eqs. (\ref{OmegaR}) and (\ref{detuning}) are shown in Figs. \ref{few_osci} (c) and (d), respectively.
The populations of the bare basis, $P_g(t)$ and $P_e(t)$, 
are shown in Fig. \ref{few_osci} (e).
Fig. \ref{few_osci} (f) shows the time-dependent transition frequency of the atom, $\omega_0(t)/(2\pi)$, given by Eq. (\ref{delta}), which changes sign. 
This is possible by varying laser intensities around a  ``light-induced level crossing'' \cite{lange}. 

\subsection{Pulse with many field oscillations}
%
%
%
%
\begin{figure}[t]
\begin{center}
\includegraphics[height=4.0cm,angle=0]{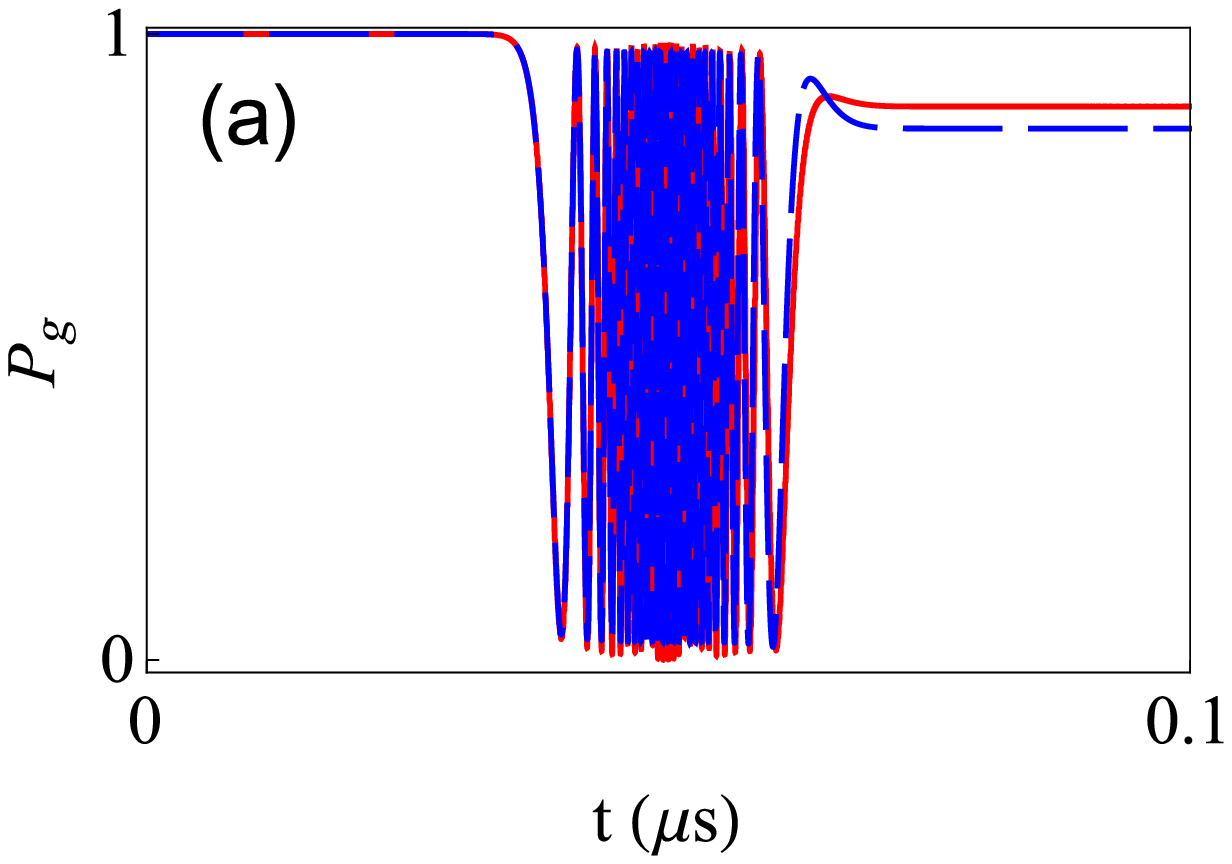}
\includegraphics[height=4.0cm,angle=0]{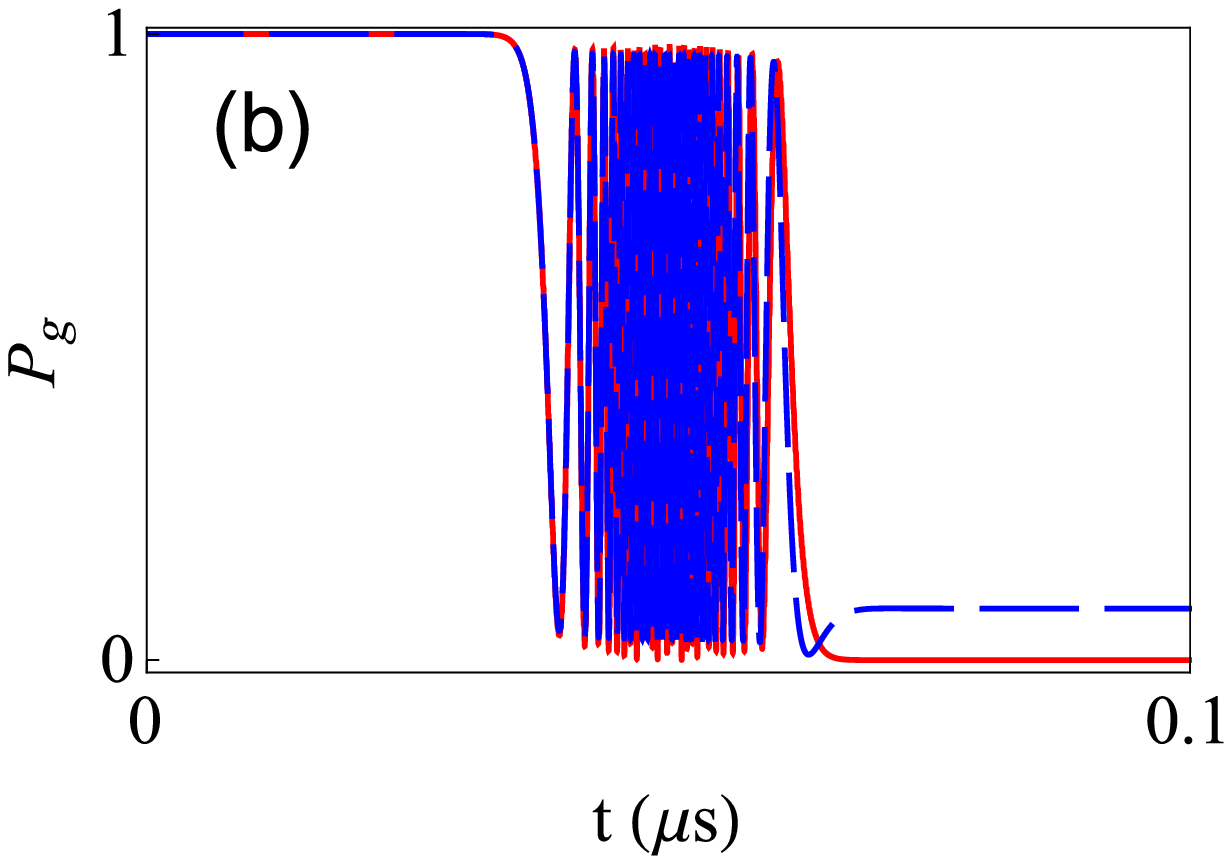}
\end{center}
\caption{\label{P1_P2}
(Color online)
$P_g$ driven by  $H_{RWA}(t)$ (blue dashed line) and by the exact Hamiltonian $H(t)$ (red solid line) for
(a) the reference parameters $a= (2\pi)^2 \times 254.648$ MHz$^2$ and
$\Omega_0= 2\pi \times 2$ GHz, and for
(b) the optimized parameters $a=  (2\pi)^2 \times 272.824$ MHz$^2$ and  
$\Omega_0= 2\pi \times 2.202$ GHz, for a population inversion process. In both cases
$t_f= 0.1$ $\mu$s, $\omega_0= 2\pi \times 5$ GHz,
and $A= (2\pi)^2 \times 506.606$ MHz$^2$.}
\end{figure}
%
%
%
For many field oscillations in the pulse the singularities to avoid in equations (\ref{OmegaR}) and (\ref{detuning})
become too numerous to apply the previous approach.  
A simple inversion method is to assume sensible specific forms with free parameters for the functions 
$\Delta$ and $\Omega_R$. 
An example is a population inversion process carried out by a linear detuning and a Gaussian Rabi frequency,
\beqa
\Delta(t) &=& a (t-t_f/2),
\label{delta_lineal} \\
\Omega_R(t) &=& \Omega_0 \exp[-A (t-t_f/2)^2],
\nonumber
\eeqa
with two free parameters, $a$ and $\Omega_0$. For a constant transition (angular) frequency, $\omega_0$, taking into account Eqs. (\ref{delta}) and (\ref{delta_lineal}), setting $\varphi(0)=\pi/2$ and integrating, the phase of the pulse is given by
\beq
\varphi(t)= -a t^2 /2 + (\omega_0+a t_f/2) t + \pi/2.
\nonumber
\eeq
Setting $\beta(0)=0$,  we may solve for $\theta(t)$ and $\beta(t)$ in the system (\ref{NRWA}), 
and minimize $[\theta(t_f)-\pi]^2$ numerically to determine $a$ and $\Omega_0$. 
This is a simple alternative to a more sophisticated optimal-control-theory approach \cite{scheuer}
or bang-bang methods \cite{avinadav}.   

In a numerical example we first set the reference parameters,
$t_f= 0.1$ $\mu$s, $\omega_0= 2\pi \times 5$ GHz, $A= (2\pi)^2 \times 506.606$ MHz$^2$,
$a= (2\pi)^2 \times 254.648$ MHz$^2$, and
$\Omega_0= 2\pi \times 2$ GHz, for which the Hamiltonian within and without the RWA give similar dynamics with unsuccessful population inversions, see Fig. \ref{P1_P2} (a).
For these values of $t_f$, $\omega_0$, and $A$, and from these seed parameters  $a$ and $\Omega_0$ a minimization algorithm provides optimized parameters for the population inversion process with the exact Hamiltonian $H(t)$,
$a=  (2\pi)^2 \times 272.824$ MHz$^2$ and $\Omega_0= 2\pi \times 2.202$ GHz. The same optimized parameters do not invert the population when the RWA is applied, see Fig. \ref{P1_P2} (b).

\section{Discussion and conclusions}
For a two-level system in a classical field  
the diagonal and non-diagonal elements in the exact interaction-picture Hamiltonian $H(t)$ must depend consistently on the phase of the field, $\varphi(t)$, and its derivative, $\dot{\varphi}(t)$. This makes the inverse engineering methods more complicated than with the Hamiltonian within the rotating-wave approximation, $H_{RWA}(t)$.
Simple attempts using invariants or counterdiabatic methods to implement faster-than-adiabatic processes may not satisfy the consistency condition in $H(t)$ or lead to singularities.  Different ways have been shown to circumvent the difficulties. While we have presented simple proof-of-principle examples, the methods may 
be adapted to minimize effects of noise and decoherence due to the flexibility of the inversion \cite{NJP}.

An open problem is to extend the  approaches to systems where more levels have to be considered.  
In many systems the failure of the RWA is associated with the need to include  further 
levels in the theoretical treatment. In trapped ions, for example, when the vibrational RWA is not applied, and vibrational counter-rotating terms are taken into account, 
the energy levels are distorted and the sideband resonances are shifted \cite{lizuain}. This may be understood as a vibrational Bloch-Siegert effect or, equivalently, as the result of Stark shifts of the levels due to off-resonant transitions \cite{lizuain}. 
Nevertheless, it is possible to describe the subspace of the two states in an  isolated anticrossing by an approximate $2\times2$ Hamiltonian  that takes into account the effect of further levels perturbatively by means of a level shift operator \cite{Cohen,lizuain}. The current approaches could then be applied but the details 
are left for a separate study.  

\acknowledgements{
We thank G. Romero for discussions.  
We acknowledge funding from Projects No. IT472-10 and No. FIS2012-36673-C03-01,
and the UPV/EHU program UFI 11/55. This work was partially supported by the NSFC (11474193, 61176118),
the Shuguang and Pujiang Program (14SU35, 13PJ1403000),  the Specialized Research Fund for the Doctoral Program (2013310811003),
and the Program for Eastern Scholar. 
S. I. acknowledges UPV/EHU for a postdoctoral position.}

\end{document}